\documentclass[twocolumn,aps,pra,prashowpacs,floatfix,notitlepage,longbibliography,superscriptaddress]{revtex4-1} 
\usepackage{calrsfs}
\usepackage[normalem]{ulem}

\usepackage{graphics,dcolumn}
\usepackage{bm}
\usepackage[fleqn]{amsmath}
\usepackage{hyperref}
\hypersetup{
     colorlinks=true,       
    linkcolor=blue,          
   citecolor=blue,        
   filecolor=magenta,      
   urlcolor=cyan           
}

\usepackage{amsfonts,amssymb}
\usepackage{graphics,dcolumn}
\usepackage{bbold}
\usepackage{amsmath}
\usepackage{graphics,dcolumn}
\usepackage{bm}
\usepackage[fleqn]{amsmath}
\usepackage{hyperref}
\hypersetup{
     colorlinks=true,       
    linkcolor=blue,          
    citecolor=blue,        
    filecolor=magenta,      
    urlcolor=cyan           
}
\usepackage{exscale,relsize}
\usepackage{epsfig}
\usepackage{amssymb}
\usepackage{amsmath}
\usepackage{euscript}
\usepackage{ulem}
\usepackage{float}
\usepackage{bbm}
\usepackage{footmisc}
 \usepackage{amsthm}

\newcommand{\be}{\begin{equation}}
\newcommand{\e}{\end{equation}}

\newcommand{\beml}{\begin{subequations}}
\newcommand{\eml}{\end{subequations}}
\newcommand{\beg}{\begin{eqnarray}}
\newcommand{\eq}{\end{eqnarray}}
\newcommand{\ba}{\begin{array}}
\newcommand{\ea}{\end{array}}

\newcommand{\lt}{\left}
\newcommand{\rt}{\right}
\newcommand{\n}{\nonumber}

\newcommand{\s}{\sigma}
\newcommand{\la}{\langle}
\newcommand{\ra}{\rangle}

\newcommand{\im}{\,{\rm Im}\,}
\newcommand{\re}{\,{\rm Re}\,}

\begin{document}
\date{\today}

\title{Coherence turned on by incoherent light}

\author{Vyacheslav N. Shatokhin}
\affiliation{Physikalisches Institut, Albert-Ludwigs-Universit\"at Freiburg, Hermann-Herder-Str. 3,
D-79104 Freiburg, Federal Republic of Germany}

\author{Mattia Walschaers}
\affiliation{Physikalisches Institut, Albert-Ludwigs-Universit\"at Freiburg, Hermann-Herder-Str. 3,
D-79104 Freiburg, Federal Republic of Germany}
\affiliation{Instituut voor Theoretische Fysica, KU Leuven, Celestijnenlaan 200D, B-3001 Heverlee, Belgium}
\affiliation{Laboratoire Kastler Brossel, UPMC-Sorbonne Universit\'es, CNRS, ENS-PSL Research University, Coll\`ege de France; 4 place Jussieu, F-75252 Paris, France}

\author{Frank Schlawin}
\affiliation{Physikalisches Institut, Albert-Ludwigs-Universit\"at Freiburg, Hermann-Herder-Str. 3,
D-79104 Freiburg, Federal Republic of Germany}
\affiliation{Clarendon Laboratory, University of Oxford, Parks Road, Oxford OX1 3PU, United Kingdom}

\author{Andreas Buchleitner}
\affiliation{Physikalisches Institut, Albert-Ludwigs-Universit\"at Freiburg, Hermann-Herder-Str. 3,
D-79104 Freiburg, Federal Republic of Germany}

\begin{abstract}
One of the most pertinent problems in the debate on non-trivial quantum effects in biology concerns natural photosynthesis. Since sunlight is composed of thermal photons, it was argued to be unable to induce quantum coherence in matter, and that quantum mechanics is therefore irrelevant for the dynamical processes following photoabsorption. Our present analysis of a toy ``molecular aggregate" -- composed of two dipole-dipole interacting two-level atoms treated as an open quantum system -- however shows that incoherent excitations indeed can trigger coherent dynamics that persist: We demonstrate that collective decay processes induced by the dipole-dipole interaction create coherent intermolecular transport -- regardless of the coherence properties of the incoming radiation. Our analysis shows that the steady state coherence  is mediated by the population imbalance between the molecules and, therefore, {\it increases} with the energy difference between the two-level atoms. Our results establish 
the importance of collective decay processes in the study of ultrafast photophysics, and especially their potential role to generate stationary coherence in incoherently driven quantum transport.
\end{abstract}

\maketitle

\section{Introduction}
A detailed understanding of the microscopic processes which underlie natural photosynthesis represents an important and intriguing source of inspiration for technologies which seek to efficiently capture, transform, and store solar energy \cite{Scholes:2011uq,blanckenship}. One of the most important open questions in this research area is whether quantum interference effects play a role in solar light harvesting, and possibly could be used for highly efficient solar energy conversion \cite{Scholes17,Duan17}. That transient quantum coherence can prevail in such complex structures, at ambient temperatures, can by now be regarded a solidly established experimental fact \cite{Brixner:2005jb,Engel:2007yq}, and has also been reported for the charge separation process in organic solar cells \cite{Andrea-Rozzi:2013uq, Falke14}. 
 
It however is argued \cite{Brumer27112012} that the evidence provided by the above experiments is inconclusive, because the conditions under which quantum effects were experimentally observed in certain light harvesting complexes (LHC) differ from conditions in vivo. Indeed, laboratory experiments rely on photon echo spectroscopy \cite{mukamel_book}, where the energy transfer is induced by a series of ultrashort coherent laser pulses. In contrast, sunlight (driving the natural process) can be described as continuous wave (or stationary) thermal (incoherent) radiation \cite{mandel95}. Thus, it is a priori crucial to distinguish the coherence observed in photon echoes \cite{pullerits12} from coherence which may arise in non-equilibrium open system quantum dynamics -- as we will outline below.
 
Moreover, some models \cite{Brumer27112012,brumer13,1367-2630-12-6-065044} suggest that the coupling of a quantum system to a thermal radiation bath rapidly leads to the formation of a stationary state that does not exhibit any coherences. 
This apparently contradicts the point of view that {\it coherent} non-equilibrium transport processes -- leading to the observed efficiency of the excitation transfer \cite{fleming09,Ishizaki12,manzano12} -- can be triggered by any photoabsorption event \cite{Dorfman19022013}, regardless of the source of photons.
 
Here we develop a microscopic quantum optical theory to resolve this longstanding controversy. Specifically, we establish that steady state coherence can indeed emerge in an incoherently driven molecular complex, under realistic assumptions on the incident wave lengths and molecular separations. To begin with, we recall that the primary process of photosynthesis is the absorption of a single photon by a chlorophyll molecule, whereby the molecule undergoes a transition from the ground to the excited electronic state \cite{blanckenship}. The photoabsorption initiates energy transfer towards the reaction center, where a charge separation cascade with almost unit efficiency is triggered \cite{struempfer12}. This transfer process from the initial absorption event to the charge separation has a finite duration, of the order of 10-100 picoseconds \cite{Dostal:2016fk}, and it is during this process that transient electronic coherences have been observed \cite{Brixner:2005jb,Engel:2007yq,Dostal:2016fk,Collini:2010nz}. Afterwards, the molecule resets in its ground electronic state and is able to absorb the next photon. We will show that, when averaging over many such single photon absorption and transfer cycles, one ends up with a master equation-type ensemble description which exhibits non-vanishing coherence in the non-equilibrium steady state. 

\begin{figure}
\includegraphics[width=8cm]{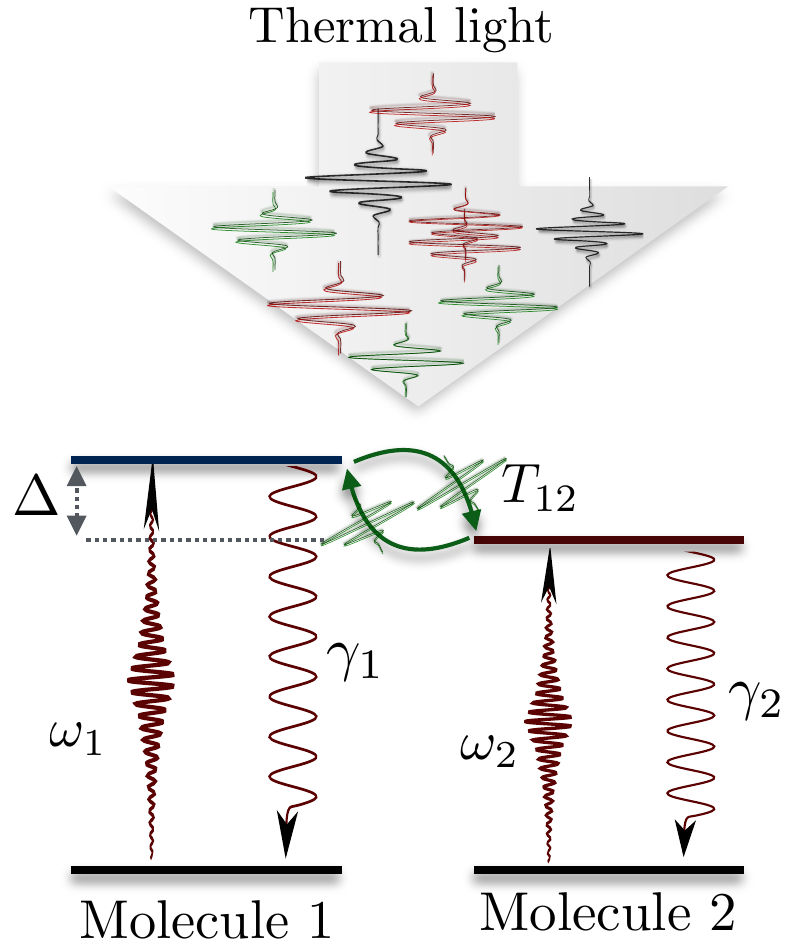}
\caption{Two molecules at a distance $r_{12}$ and with transition frequencies $\omega_1$ and $\omega_2$, 
detuned 
by $\Delta=\omega_1-\omega_2$, are embedded into a common electromagnetic bath. The bath induces radiative decay of the molecules ($\gamma_k$ is the decay rate of molecule $k$), as well as a dipole-dipole interaction with complex coupling constant $T_{12}$ [see Eq.~(\ref{fom})]. 
An external, incoherent thermal source stimulates 
absorption and 
emission processes. }
\label{schema_old}
\end{figure}

We model an LHC as a ``molecular aggregate" which consists of two effective two-level atoms  \cite{MayKuehn04} -- which we shall refer to as molecules in the following -- that are embedded into a common electromagnetic bath. Thereby, we abstract ourselves from the details of the structure and energy spectra of a real photosynthetic complex \cite{blanckenship}. Nonetheless,  our dimer model is able to describe two absorption bands associated with the widths of the electronic excited states, as well as the dipole-dipole interaction between the molecules.
We study the interaction of this system with an external incoherent field which represents the sunlight, and show that 
coherent evolution
 survives 
 even 
 in the non-equilibrium {\it steady} state of the incoherently driven system, as a reflection of the transient coherences induced on the level of single photon absorption and transport processes.

\section{Model}
Our model is presented in Fig.~\ref{schema_old}. 
It consists of two molecules embedded, at a distance $r_{12}$, into a common radiation bath and interacting with an external incoherent radiation field in the optical frequency range. We assume that the molecules have allowed dipole transitions between their electronic ground and excited states $|g_k\rangle$ and $|e_k\rangle$, $k=1,2$, respectively, and that their optical transition frequencies $\omega_1$ and $\omega_2$ are detuned with $\Delta=\omega_1-\omega_2 \ll \omega_1,\omega_2$.
Furthermore, we can ignore the ambient thermal photons at optical frequencies and therefore assume the relevant modes of the radiation reservoir in the vacuum state. This bath induces spontaneous decay of the individual molecules with rates $\gamma_k$, as well as their dipole-dipole interaction with complex coupling strength $T_{12}$. Additionally, the coupling to other, e.g. vibrational, degrees of freedom may cause further dissipation \cite{fleming09,Renger2001137}, which is not considered in this work.
As for the external incoherent field, we assume that its energy density is a slowly varying function around the transition frequency. The external field generates absorption and stimulated emission processes at the rate $\gamma_kN(\omega_k)$ \cite{loudon}, where $N(\omega_k)$ is the average number of the (incoherent) source photons at the transition frequency, which is defined by the source temperature.

Using standard quantum optical methods \cite{scully,agarwal74}, one can trace out the bath degrees of freedom, to arrive at the master equation governing the evolution of the ``aggregate'' density  matrix $\rho$, in the basis of the uncoupled individual molecules' energy eigenstates $\{|g_1,g_2\ra, |e_1,g_2\ra, |g_1,e_2\ra, |e_1,e_2\ra\}$.  In the frame rotating at the 
average
frequency $\omega_0=(\omega_1+\omega_2)/2$ 
wherein rapidly oscillating terms are eliminated since $\omega_1\approx \omega_2$,  
the master equation reads
\begin{widetext}
\begin{align}
\dot{\varrho}&=\sum_{k\neq l=1}^2\lt( \frac{i\Delta}{2}(-1)^{k}[\sigma_+^k\sigma_-^k,\varrho]-i\Omega[\sigma_+^k\sigma_-^l,\varrho]
+\gamma_kN(\omega_k)\{[\sigma_+^k,\varrho\sigma_-^k]+[\sigma_+^k\varrho,\sigma_-^k]\}\rt.\n\\
&\lt.+\gamma_k\{1+N(\omega_k)\}\{[\sigma_-^k,\varrho\sigma_+^k]+[\sigma_-^k\varrho,\sigma_+^k]\}
+\Gamma\{[\sigma_-^k,\varrho\sigma_+^l]+[\sigma_-^k\varrho,\sigma_+^l]\}\rt).
\label{meq}
\end{align}
\end{widetext}
In this equation, the atomic (de-)excitation operators are given as $\sigma_-^k=|g_k\ra\la e_k|$, $\sigma_+^k=|e_k\ra\la g_k|$, the atomic decay rates read
$\gamma_k=d_k^2\omega_k^3/6\pi\epsilon_0\hbar c^3$, and $\Gamma\equiv\Gamma(\omega_0r_{12}/c)$, $\Omega\equiv \Omega(\omega_0r_{12}/c)$ are the real and imaginary parts, respectively, of the {\it retarded} dipole-dipole interaction strength $T_{12}=\Gamma+i\Omega$, which in particular generates collective effects such as super-radiance \cite{agarwal74,Andrews1987,salama}\footnote{In the chemical physics literature \cite{Andrews1987,salama,frost2014}, the
  complex retarded dipole-dipole interaction strength is defined as $\Omega
  +i\Gamma $, which up to a phase factor coincides with the one adopted in this
  work.}. The physical meaning of the real and imaginary parts of $T_{12}$ can be unambiguously 
identified from the structure of the master equation (\ref{meq}): 
Terms proportional to $i\Omega$ describe oscillatory, reversible, non-radiative excitation exchange between both molecules, and lead to the formation of delocalized excitonic states. Terms proportional to 
$\Gamma$ represent (collective) radiative decay processes, following a non-radiative excitation exchange between the molecules. Accordingly,
$\Gamma$ and $\Omega$  are associated with the life time and the energy shift of the (entangled, Dicke) eigenstates $|\psi_+\ra$, $|\psi_-\ra$ (see Appendix \ref{app1}) of the dipole-coupled molecular dimer, respectively \cite{agarwal74,Gross1982301,ficek02}. Explicitly, $\Gamma$ and $\Omega$ are given as \cite{agarwal74,Andrews1987,lehmberg70},
\begin{widetext}
\begin{subequations}
\begin{align}
\Gamma(\xi)&\equiv \frac{3\sqrt{\gamma_1\gamma_2}}{2}\left\{[\hat{\bf d}_1\cdot\hat{\bf d}_2\!-\!(\hat{\bf d}_1\cdot\hat{\bf r}_{12})(\hat{\bf d}_2\cdot\hat{\bf r}_{12})]\frac{\sin\xi}{\xi}
+[\hat{\bf d}_1\cdot\hat{\bf d}_2\!-\!3(\hat{\bf d}_1\cdot\hat{\bf r}_{12})(\hat{\bf d}_2\cdot\hat{\bf r}_{12})]\right.\nonumber\\
&\left.\times\left(\frac{\cos\xi}{\xi^2}\!-\!\frac{\sin\xi}{\xi^3}\right)\right\},\label{f}\\
\Omega(\xi)&\equiv \frac{3\sqrt{\gamma_1\gamma_2}}{2}\left\{\!-\![\hat{\bf d}_1\!\cdot\!\hat{\bf d}_2\!-\!(\hat{\bf d}_1\!\cdot\!\hat{\bf r}_{12})(\hat{\bf d}_2\!\cdot\!\hat{\bf r}_{12})]\frac{\cos\xi}{\xi}
+[\hat{\bf d}_1\cdot\hat{\bf d}_2\!-\!3(\hat{\bf d}_1\cdot\hat{\bf r}_{12})(\hat{\bf d}_2\cdot\hat{\bf r}_{12})]\right.\nonumber\\
&\left.\times\left(\frac{\sin\xi}{\xi^2}+\frac{\cos\xi}{\xi^3}\right)\right\},\label{om}
\end{align}
\label{fom}
\end{subequations}
\end{widetext}
where $\xi\equiv \omega_0r_{12}/c$ is the effective intermolecular distance\footnote{For optical wave lengths of $400-900$ nm, and for typical distances $r_{12}$
  between the chlorophyll molecules in different LHCs varying from 1 to 10 nm
  \cite{blanckenship,milder10}, effective distances lie in the range $\xi \sim
  0.01-0.1$.}, and $\hat{\bf d}_k$ and $\hat{\bf r}_{12}$ are unit vectors directed along the $k$th  molecular dipole and along the vector connecting the molecules, respectively. Note that the far-field terms in Eqs. (\ref{f},\ref{om})  (i.e., the terms decreasing as $\xi^{-1}$ for $\xi\gg 1$) describe retardation effects proper that are associated with the exchange of real photons \cite{Andrews1987}. These effects start playing a role at $r_{12}\gtrsim 10$ nm \cite{frost2014}, though they are deemed unimportant at inter-molecular distances 
of less than 10 nm (i.e. $\xi\lesssim 0.1$).

\begin{figure*}
{\includegraphics[width=18cm]{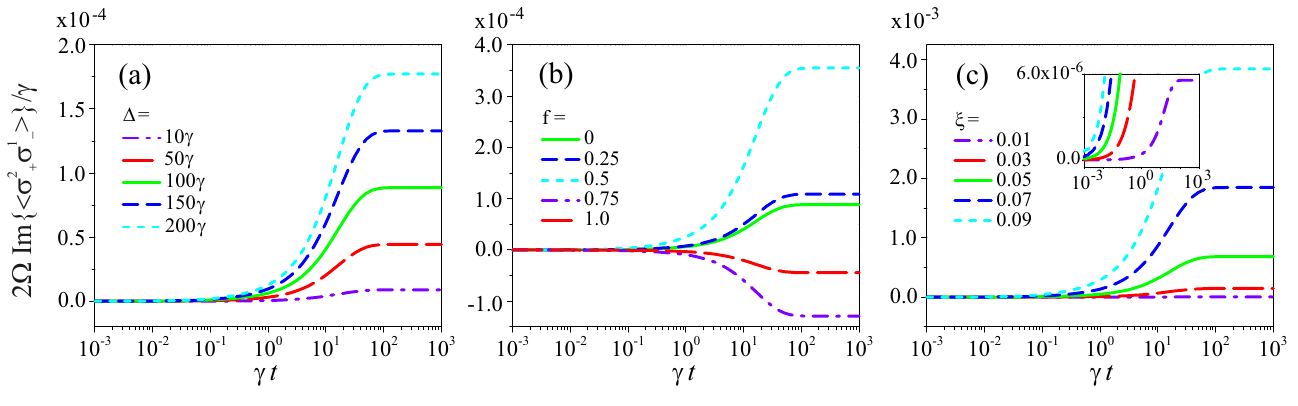}}
\caption{
The inter-molecular electronic-coherence-induced excitation current $2\Omega{\rm Im}\{\la \sigma_+^2(t)\sigma_-^1(t)\ra\}$, as generated by Eq.~(\ref{meq}) (on a semi-log scale), with $N(\omega_1)=N(\omega_2)=N(\omega_0)=0.01$ (corresponding to a
source temperature $T\approx 6000$ K), for realistic choices of the
characteristic parameters (see legends) which define the molecular dimer's dynamics:
(a) Variation of the molecular detuning $\Delta$,  at $\xi=0.02$  (or $r_{12}=1.5\ \rm nm$), ${\rm f}=0$. (b) Variable dipole orientation ${\rm f}$,  
at $\xi=0.02$,  $\Delta=100\gamma$. (c) Variable distance $\xi$ (the values 0.01, 0.03, 0.05, 0.07, 0.09 correspond to $r_{12}=0.73, 2.2, 3.7, 5.1, 6.6\ \rm nm$, 
respectively) of the dimer's constituent molecules, 
at ${\rm f}=0$, $\Delta=50\gamma$.}
\label{model_a}
\end{figure*}

One of the key processes in the theory of photosynthesis is resonance energy transfer \cite{MayKuehn04}. This transfer is effective between molecules whose transition frequencies are close to each other (hence, the name of the process) and is characterized by a rate proportional to $|\Gamma+i\Omega|^2$ \cite{Andrews1987}. In the non-retarded limit $\xi\ll 1$, $\Gamma (\xi)$ is much smaller than $\Omega(\xi)$. It is therefore common practice to neglect $\Gamma(\xi)$, and to retain only the non-retarded contributions of $\Omega(\xi)$ \cite{MayKuehn04,PhysRevE.83.021912,cao2014}.
In this limit, $\Omega(\xi)\to V_{dd}/\hbar$, with $V_{dd}$ the static 
dipole-dipole interaction energy 
\cite{agarwal74,salama}:
\begin{equation}
V_{dd}=\frac{{\bf d}_1\cdot{\bf d}_2-3({\bf d}_1\cdot\hat{\bf r}_{12})({\bf d}_2\cdot\hat{\bf r}_{12})}{4\pi\epsilon_0 r_{12}^3}\ .
\label{Vdd}
\end{equation}
This wide-spread approximation however neglects that 
also 
$\Gamma(\xi)$ does remain finite as $r_{12}\to 0$, with $\Gamma (\xi) \rightarrow  \sqrt{\gamma_1\gamma_2}\hat{\bf d}_1\cdot\hat{\bf d}_2$,
and Eq.~(\ref{Vdd}) is thus imprecise at small distances.
As we show below, a consequence of
using the approximate expression (\ref{Vdd}) is that a collective coherent effect Ð-- the stationary excitation
current in the dipole-interacting system Ð-- is erroneously predicted to vanish.

Let us inspect the time-dependent expectation value $\im\{\la\sigma_+^2(t)\sigma_-^1(t)\ra\}\equiv \im\{\la e_1,g_2
|\varrho(t)|g_1,e_2\ra\}$ as the quantifier of the electronic inter-site coherence of our ``molecular aggregate'' under incoherent driving. Upon multiplication by $2\Omega$  this yields the {\it excitation current}, which is proportional to the probability per unit time for an excitation transfer from molecule 1 to molecule 2 (see Appendix \ref{sec:app}). For simplicity and without loss of essential physics, we study the temporal behavior of this coherence-induced current under the assumptions that the thermal source is characterised by $N(\omega_0)=0.01$ (which is consistent with the mean photon number of the sunlight at the optical frequencies), both dipoles point in the same direction, and that the excited states of molecules 1 and 2 have equal linewidths  $\gamma_1=\gamma_2=\gamma$\footnote{This can be justified by using a standard expression for the spontaneous decay rate $\gamma_k$ [see its definition following Eq. (\ref{meq})]. If we assume that both dipoles have equal matrix elements and their transition frequencies lie in the optical domain, then for $\omega_1=2\pi\times 10^{15}$ Hz, $\omega_2=\omega_1-\Delta$ and $\Delta= 10$ GHz (which corresponds to $\simeq 100$ natural linewidths), we obtain, $\gamma_2/\gamma_1\approx 1-3\Delta/\omega_1=0.99997$.}. Furthermore, we assume that $\Delta>0$, by noting that in a fully symmetric system, where $\gamma_1=\gamma_2$ and $\Delta=0$, the expectation value of the excitation current trivially vanishes for all times, i.e., $\im\{\la\sigma_+^2(t)\sigma_-^1(t)\ra\}\equiv 0$, while the treatment of the case $\Delta<0$ amounts to relabelling molecules 1 and 2.  

Our results on the temporal evolution of $2\Omega\im\{\la\sigma_+^2(t)\sigma_-^1(t)\ra\}$ are plotted in Fig.~\ref{model_a}(a)-(c), 
where we vary the detuning $\Delta$, the orientation ${\rm f}\!=\!(\hat{\bf d}_1\!\cdot\!\hat{\bf r}_{12})$, or the effective distance $\xi$, respectively, while keeping the two remaining parameters fixed. It is evident that a non-vanishing
current is a generic feature of the intramolecular excitation transfer that follows the photo-absorption process 
by the molecular aggregate
prepared in its ground (reset) state at $t=0$. 
The non-equilibrium coherence emerges on time scales $t \gtrsim 10^{-3}\gamma^{-1}$, when radiative relaxation processes come into play. At  $t\gg\gamma^{-1}$, the excitation current tends to its steady state value mono-exponentially,
\begin{align}
2\Omega\im\{\la\sigma_+^2(t)\sigma_-^1(t)\ra\}\!&=\!\frac{4N(\omega_0)\Omega\Gamma[1\!+\!2N(\omega_0)]\gamma^2\Delta\!}{R}\nonumber\\
&\times\{1-\exp(-C t)\}
 \label{exc_cur0},
\end{align}
where $C\equiv C(\xi,{\rm f},\Delta)\sim \gamma N(\omega_0)$ and $R$ is given in Eq. (\ref{R}).

Equation (\ref{exc_cur0}) shows that the stationary excitation current {\it only}  emerges for a non-vanishing collective decay rate 
$\Gamma$, giving 
rise to the irreversibility of the excitation exchange process \footnote{Finite steady state electronic coherence may arise for vanishing $\Gamma $ if
one allows for the coupling of the electronic excitations to additional
degrees of freedom, as discussed in \cite{cao2014}.}. As a result, the stationary populations of the excited levels of the two molecules become unequal, 
which
is crucial for the emergence of the stationary current. 
It is also evident from Eq. (\ref{exc_cur0}) that 
the population imbalance and, hence, 
the current in the steady state {\it increase} 
with $\Delta$. 
More explicitly, the stationary coherence and the population imbalance are in fact proportional to one another (see Appendix \ref{sec:app}):
 \be 
 2\Omega \im\{\la\sigma_+^2\sigma_-^1\ra\}\!=\!\frac{\kappa}{2}\{\la \sigma_+^2\sigma_-^2\ra\!-\!\la \sigma_+^1\sigma_-^1\ra\}\ .\label{en_bal}
\e
Here, $\kappa=2\gamma(1+2N)$ and $\la\sigma_+^k\sigma_-^k\ra=\la e_k|\varrho^k|e_k\ra$, with $\varrho^k={\rm Tr}_l\!(\varrho)$ ($k,l=1,2$, $k\neq l$) the reduced density matrix of molecule $k$.
This result can be interpreted as an energy balance relation for our dipole-dipole coupled system: The left hand side of (\ref{en_bal}) yields the number of photons that is transferred per unit time from molecule 1 to molecule 2; the right hand side yields the difference between the total number of photons that are emitted, or absorbed, per unit time, by molecule 2 and 
1, due to spontaneous and stimulated emission. 

\begin{figure*}
{\includegraphics[width=14.0cm]{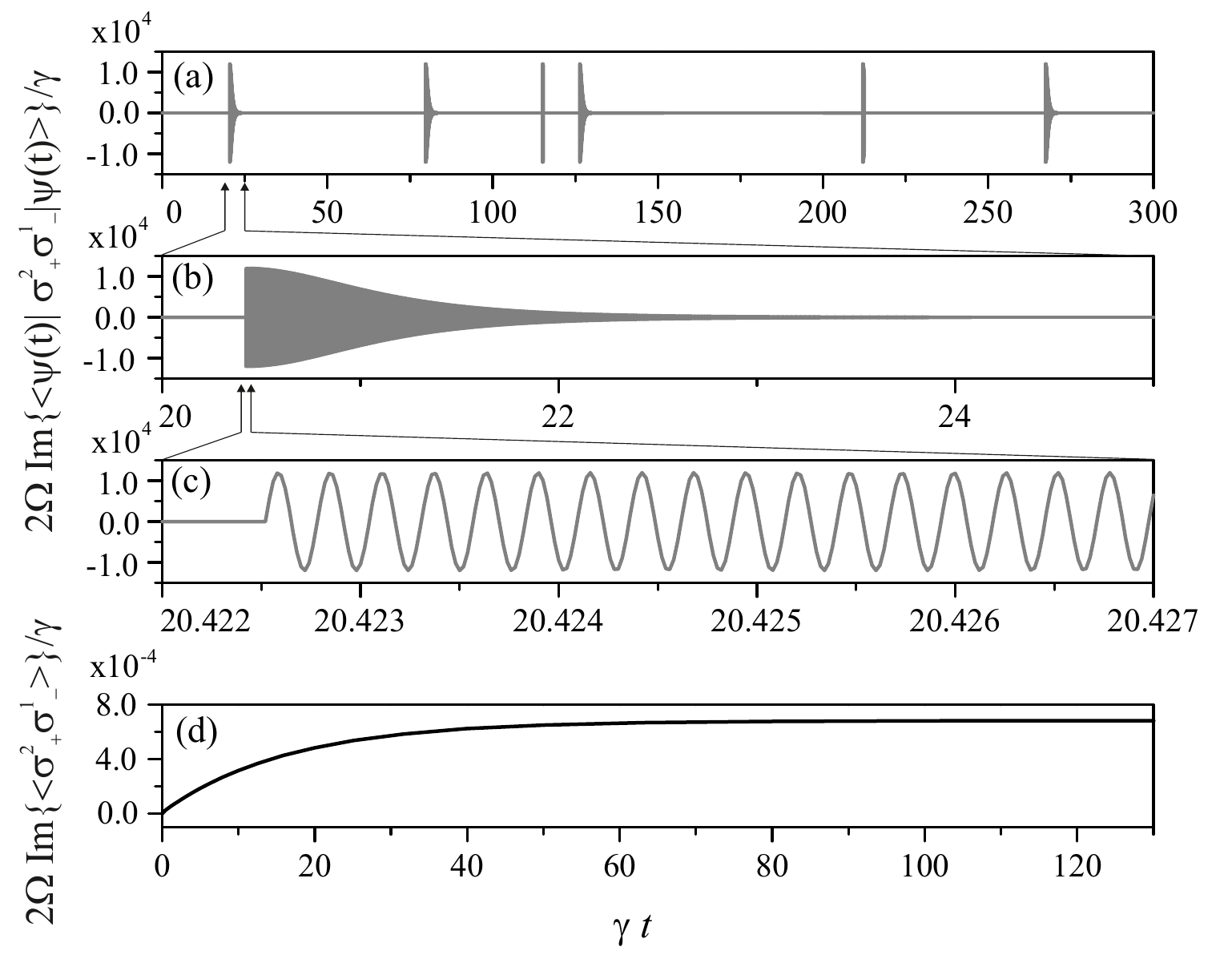}}
\caption{
The intermolecular excitation current at $\xi=0.05$, $f=0$, and $\Delta=50\gamma$ obtained as (a-c) quantum mechanical expectation value with respect to the ``quantum trajectory'' $|\psi(t)\ra$, where (a) displays the first six coherent transients of the current initiated (interrupted) by photon absorption (emission) events, (b) zooms in the first coherent transient 
which tends to a finite value (indiscernible in the plot) at $\gamma (t-t_0)\gtrsim 10\gamma^{-1}$ before the next jump occurs, 
and (c) zooms in a sequence of coherent oscillations of the current following the first thermal photon absorption. The initially (right after photon absorption) positive current indicates that the photon has been absorbed by molecule 2.
(d) Ensemble average over random realizations of $|\psi(t)\rangle$ yields the  intermolecular excitation current $2\Omega{\rm Im}\{\langle\sigma^2_+\sigma_-^1\rangle\}/\gamma$. Note the different scale of the $y-$axis in (d) as compared to (a-c).
}
\label{traj}
\end{figure*}

Equation (\ref{en_bal}) is reminiscent of relations well known in single atom resonance fluorescence \cite{cohen_tannoudji_API,shatokhin02}: There, quantum coherence between the atomic ground and excited states arises due to the presence of the laser field, characterized by the Rabi frequency. The quantity describing the atomic coherence, $\im\{\la \sigma_+\ra\}$, is coupled to the atomic excited state population, such that single atom energy balance relations similar to (\ref{en_bal}) hold. In our present case of two molecules, quantum coherence between the molecular dipoles arises due to the  dipole-dipole interaction, playing the role of the Rabi frequency. The directed excitation current leads to an imbalance of the molecular excited state populations: the molecule with larger transition frequency $\omega_1>\omega_2$ has smaller excited state population because part of it is coherently transferred to the molecule with smaller transition frequency $\omega_2$. The magnitude of the current is highly sensitive to the dimer's parameters and is typically $\simeq 10^2-10^4$ excitation transfer events per second (see Fig. \ref{model_a}), or about $0.01-1$ \% of the incoherent pumping rate $\gamma N(\omega_0)\simeq 10^6$ s$^{-1}$, assuming $\gamma\simeq 10^8$ s$^{-1}$. 

The above scenario of the downhill excitation current is apparently violated for the orientations ${\rm f}$ for which $\Omega$ is negative (see, e.g., the dashed-dotted and long dashed lines in Fig. \ref{model_a}(b)). Yet, in the dimer's eigenbasis the stationary current always flows towards the state with smaller energy (see Appendix \ref{app1}).

\section{Unravelled dynamics}
At first glance, the monotonic build-up of the  stationary current in Fig.~\ref{model_a} may seem inconsistent with the transient character of the observed 
quantum coherences \cite{Brixner:2005jb,Engel:2007yq}. However, the typical behavior of the current can be viewed as a result of an average over an ensemble of ``quantum trajectories'' \cite{carmichael,moelmer93} (see Fig. \ref{traj}(a)) corresponding to individual incidents of an excitation process in
our ``molecular aggregate" (see Appendix \ref{sec:mc}): Initially (re-)set in their ground
states, molecule 1 or 2 absorbs a photon (undergoes a ``quantum jump'' mediated by the operator $\sigma_+^1$ or $\sigma_+^2$), with relative 
probability one half, at a random moment in time $t_0 > 0$. 
The photoabsorption 
prepares the dimer in either the state $|e_1,g_2\ra$ or $|g_1,e_2\ra$ and launches coherent evolution within the single excitation subspace governed by the Hamiltonian $H_D$ (see Appendix \ref{app1}). 
The coherent exchange of the excitation between the molecules generated by the latter Hamiltonian translates    
into a transient oscillation of the excitation current at frequency $\sqrt{4\Omega^2+\Delta^2}$, see Fig. \ref{traj}(c). 
If a photon is emitted by the dimer at a random time $t_1$ such that $t_1-t_0<\gamma^{-1}$, then this  
happens primarily as a result of a quantum jump described by the collective operator $(\sigma^1_-+ \sigma^2_-)$ (see Appendix \ref{sec:mc}). The dimer is then reset to its ground state, until the next photon absorption occurs. 
For some incidents of the excitation process, the photoemission does not occur at times $t_1-t_0<\gamma^{-1}$ and   
the envelope of the current exponentially decreases on a time scale $\sim 10\gamma^{-1}$ (Fig. \ref{traj}(b)) to a finite value (see Appendix \ref{sec:mc}).  
This corresponds to the continuous evolution of the dimer into a conditioned state that is close to the (long-lived) subradiant state $|\psi(t-t_0)\ra\propto (|e_1,g_2\ra-|g_1,e_2\ra)$. From the latter state, the dimer can undergo a quantum jump into the ground state described by the collective operator $(\sigma^1_-- \sigma^2_-)$ and emit a photon or, with a smaller probability, into the doubly excited state $|e_1,e_2\ra$ by absorbing the next photon. The dimer in the state $|e_1,e_2\ra$ rapidly (on a timescale $< \gamma^{-1}$) decays into the state $|g_1,g_2\ra$ via two subsequent quantum jumps mediated by the collective operator $\sigma^1_-+\sigma^2_-$ and accompanied by the emission of two photons: the first jump brings the dimer in the superradiant state $\propto(|g_1,e_2\ra+|e_1,g_2\ra)$ while the second one, after a short delay, into the ground state. 

Summing over many such ``quantum trajectories'' of the random excitation current leads to the time evolution depicted in Fig. \ref{traj}(d), finally settling in the non-equilibrium steady state. The latter state can also be obtained as the time average over a single quantum trajectory (see Appendix \ref{sec:mc}). It is therefore not surprising that averaging the oscillatory 
current over time yields a steady state value that is eight orders of magnitude smaller [compare Fig.\ref{traj}(a-c) and (d)]. 
However, only in an asymmetric dimer system ($\Delta\neq 0$) this value is strictly distinct from zero, resulting in a directed excitation flow.  \\

\section{Conclusion}

We have studied the dynamics of the electronic coherence of a toy ``molecular aggregate'' composed of two closely located two-level atoms coupled to a vacuum reservoir and excited by an incoherent field. 
This model, both in the ensemble average as well as on the level of single quantum trajectories, accounts for the single-photon excitation process which is crucial \cite{brumer17,whaley18} for understanding the dynamics of the energy transfer in light-harvesting systems.
We have shown that, following the photoabsorption by either of the molecules, the transient behavior of this 
quantity exhibits coherent oscillations which are indicative of the excitation exchange between the dimer's constituents. The amplitude, frequency and decay rate of these oscillations are defined by the 
inter-molecular dipole-dipole interaction strength and by the local relaxation rates of the individual molecular sites' excitations. Furthermore, we have established the emergence of {\it stationary} coherence in the non-equilibrium steady state of the aggregate, giving rise to a stationary current, as a 
consequence of dipole interaction-induced collective decay processes which prevail at small inter-molecular distances, despite being usually associated  with the retarded limit. When neglected in a non-retarded theoretical description of the system \cite{MayKuehn04}, the incoherent excitation instantaneously 
creates an incoherent mixture of eigenstates, and 
steady state coherence is absent.
Thus, in contrast to the results of \cite{Brumer27112012}, our results establish a realistic scenario where intermolecular electronic steady state coherence can be triggered by the absorption of photons coming from an incoherent source, mediating transient population oscillations which relax into a coherent and directed flux of excitations in the steady state. In the future, it will be interesting to look at the interplay between light-mediated coherence and the vibrational degrees of freedom, which could give rise to unexpected effects.

Finally, the emergence of the excitation current studied here is somewhat akin to the directed flow of electrons \cite{anderson}, phonons \cite{kilin98} or atoms \cite{ponomarev06} 
in presence of relaxation, but, unlike the latter examples, features a coherent transfer process driven by collective decay. Thus, the predicted effect defines a hitherto ignored potential resource for generating quantum transport.

\begin{acknowledgments}
The authors are grateful to Graham Fleming and Greg Scholes for stimulating discussions.  
A.B. thanks Dieter Jaksch and Keble College for their generous hospitality in spring 2016. A.B. and V.N.S. 
acknowledge support through the EU Collaborative project QuProCS (Grant Agreement 641277). 
M.W. would like to thank the German National Academic Foundation for financial support. F.S. acknowledges financial support of the European Research Council under the European Union's Seventh Framework Programme (FP7/2007-2013)/ERC Grant Agreement No. 319286 Q-MAC. The authors acknowledge support by the state of Baden-W\"urtemberg through bwHPC.
\end{acknowledgments}

\appendix
\section{Eigensystem of the dimer}
\label{app1}
The first two terms of the master equation \eqref{meq} correspond to the Hamiltonian of the dimer $H_{D}=\hbar\sum_{k\neq l=1}^2\lt(\frac{\Delta}{2}(-1)^{k-1}\sigma_+^k\sigma_-^k+\Omega \sigma_+^k\sigma_-^l\rt)$, where the dipole-dipole interaction couples the states $|e_1,g_2\ra$ and $|g_1,e_2\ra$ of the non-interacting molecules. The diagonalization of  $H_{D}$ in the latter subspace yields the Dicke eigensystem, with the eigenvalues $\lambda_\pm=\pm\hbar(4\Omega^2+\Delta^2)^{1/2}/2\equiv \pm\hbar\Omega^\prime/2$ and the corresponding eigenvectors,
\beml
\begin{align}
|\psi_+\ra&=(\cos\theta|e_1,g_2\ra+\sin\theta|g_1,e_2\ra),\\
 |\psi_-\ra&=(-\sin\theta|e_1,g_2\ra+\cos\theta|g_1,e_2\ra),
\end{align} 
\label{states_s_a1}
\eml
for $\Omega>0$, and 
\beml
\begin{align}
|\psi_+\ra&=(-\cos\theta|e_1,g_2\ra+\sin\theta|g_1,e_2\ra),\\
 |\psi_-\ra&=(\sin\theta|e_1,g_2\ra+\cos\theta|g_1,e_2\ra),
\end{align} 
\label{states_s_a2}
\eml
for $\Omega<0$, where $\theta=\arctan(2|\Omega|/\Delta)/2$ and we assume $\Delta>0$. It is easy to check the two equalities that hold for the states in Eq. (\ref{states_s_a1}) and Eq. (\ref{states_s_a2}):
\be
|\psi_-\ra\la\psi_+|-|\psi_+\ra\la\psi_-|=
\begin{cases}
\sigma^+_2\sigma^-_1-\sigma^+_1\sigma^-_2, \quad \Omega>0,\\
\sigma^+_1\sigma^-_2-\sigma^+_2\sigma^-_1, \quad \Omega<0.
\end{cases}
\label{uphill}
\e
By performing the quantum mechanical average and multiplying both sides of Eq.~\eqref{uphill} by $2|\Omega|$, we obtain
\be
2|\Omega|\im\{\la|\psi_-\ra\la\psi_+|\ra\}=2|\Omega|
\begin{cases}
\im\{\la\sigma^+_2\sigma^-_1\ra\},\quad  \Omega>0,\\
\im\{\la\sigma^+_1\sigma^-_2\ra\},\quad  \Omega<0.
\end{cases}
\label{curr_loc_eig}
\e
The first case ($\Omega>0$) implies that if the current is downhill in the uncoupled basis (i.e., from $|e_1,g_2\ra$ to $|g_1,e_2\ra$) it is also downhill in the eigenbasis (i.e., from $|\psi_+\ra$ to $|\psi_-\ra$), whereas the second case means that the uphill current (from $|g_1,e_2\ra$ to $|e_1,g_2\ra$) in the uncoupled basis nevertheless corresponds to the downhill current in the eigenbasis. In either scenario, the absolute value of the current in the local basis and in the eigenbasis coincide.

\section{Solution of the master equation (\ref{meq})}
\label{sec:app}
The master equation Eq. (\ref{meq}) is equivalent to a closed linear system of 15 equations of motion for the expectation values of the 
(individual and collective) molecular operators. It is convenient to represent these 
values as elements of a vector $\la\vec{Q}\ra$:
\begin{align}
\la\vec{Q}\ra&=(\la\s^1_-\ra,\la\s^1_+\ra,\la\s^1_z\ra,\la\s^2_-\ra,\la\s^2_+\ra,\la\s^2_z\ra,\la\s^1_-\s^2_-\ra,\n\\
&\la\s^1_-\s^2_+\ra,\la\s^1_-\s^2_z\ra,\la\s^1_+\s^2_-\ra,\la\s^1_+\s^2_+\ra,\la\s^1_+\s^2_z\ra,\n\\
& \la\s^1_z\s^2_-\ra,\la\s^1_z\s^2_+\ra,\la\s^1_z\s^2_z\ra)^T,
\label{Q}
\end{align}
 where $\sigma_z^k=|e_k\ra\la e_k|-|g_k\ra\la g_k|$, and for an arbitrary operator $O$, $\la O\ra={\rm Tr}(O\varrho)$. There is a unique relation between the expectation values in Eq. (\ref{Q}) and the density matrix elements. For instance, $\la\s^1_-\ra={\rm Tr}(|e_1\ra\la g_1|\varrho)=\la g_1|\varrho^1|e_1\ra$, where $\varrho^1={\rm Tr}_2(\varrho)$ is the reduced density matrix of molecule 1, which is obtained upon tracing
over the states of molecule 2.
The resulting system of equations splits into four uncoupled subsystems. For reference, all entries of the vector $\la\vec{Q}\ra$ are listed below according to these subsystems:
\beml
\begin{align}
\text{(i)} \quad &\la\s^1_-\s^2_-\ra, \la\s^1_+\s^2_+\ra,\n\\
\text{(ii)} \quad &\la\s^1_-\ra, \la\s^2_-\ra,  \la\s^1_-\s^2_z\ra, \la\s^1_z\s^2_-\ra,\n\\
\text{(iii)} \quad &\la\s^1_+\ra, \la\s^2_+\ra, \la\s^1_+\s^2_z\ra, \la\s^1_z\s^2_+\ra,\n\\
\text{(iv)} \quad &\la\s^1_z\ra, \la\s^2_z\ra, \la\s^2_+\s^1_-\ra, \la\s^1_+\s^2_-\ra,  \la\s^1_z\s^2_z\ra.\n\label{iv}
\end{align}
\label{4lines}
\eml
The variables that are here relevant for us are contained in group (iv). Indeed, the excitation current can be expressed as a difference between the number of excitations transferred per unit time  from molecule 1 to molecule 2, minus the number of excitations that are transferred in the opposite direction, i.e. it is proportional to the two-molecule coherence function,
\be 
\frac{1}{2i}(\la\s^2_+\s^1_-\ra-\la\s^1_+\s^2_-\ra)=\im \{\la\s^2_+\s^1_-\ra\}.
\e

The latter quantity can be 
inferred from solutions of the following equation of motion:
\be
\dot{\vec{x}}=A\vec{x}+\vec{L},
\label{xt}
\e
with  $\vec{x}=(\la\s^1_z\ra, \la\s^2_z\ra, \la\s^2_+\s^1_-\ra, \la\s^1_+\s^2_-\ra,  \la\s^1_z\s^2_z\ra)^T$, 
\be A\!=\!\left(\begin{array}{ccccc}
-\kappa_1&0&-2T^*&-2T&0\\
0&-\kappa_2&-2T&-2T^*&0\\
\frac{T}{2}&\frac{T^*}{2}&-\frac{\kappa_1+\kappa_2}{2}\!-\!i\Delta&0&\Gamma\\
\frac{T^*}{2}&\frac{T}{2}&0&-\frac{\kappa_1+\kappa_2}{2}\!+\!i\Delta&\Gamma\\
-2\gamma_2&-2\gamma_1&4\Gamma&4\Gamma&-\kappa_1\!-\!\kappa_2
\end{array}
\right),
\label{A}
\e
\be
\vec{L}=(-2\gamma_1,-2\gamma_2,0,0,0)^T,
\label{L}
\e
where $\kappa_i=2\gamma_i\{1+2N(\omega_0)\}$, and $T=\Gamma+i\Omega$, with $\Gamma\equiv \Gamma(\xi)$, $\Omega\equiv \Omega(\xi)$ given by (\ref{f}) and (\ref{om}), respectively. We assume that at time $t=0$ both molecules are in their ground states, hence the vector of initial conditions is
\be
\vec{x}(0)=(-1,-1,0,0,1)^T.
\label{in_cond}
\e
 The formal time dependent solution of Eq. (\ref{xt}) reads
 \be
 \vec{x}(t)=e^{At}\vec{x}(0)+(e^{At}-\mathbb{1})A^{-1}\vec{L}.
 \label{solxt}
 \e
For arbitrary times, the temporal behavior of $\vec{x}(t)$ can be studied numerically and in Fig. \ref{model_a} we present exemplary evolutions of the excitation current $2\Omega {\rm Im}\{x_3(t)\}$. This quantity exhibits monotonic behavior, wherein the current exponentially tends to its stationary value. 

Let us now address this limit, where analytical solutions $\vec{x}(\infty)=-A^{-1}\vec{L}$ are readily available.
 
First, let us consider the steady state solutions for $\Gamma=0$. In this case, the entries of the vector $\vec{x}(\infty)$ read:
\begin{align}
\la\s^1_z\ra&=\la\s^2_z\ra=-\frac{1}{1+2N}, \\
\la\s^2_+\s^1_-\ra&=\la\s^1_+\s^2_-\ra=0, \\
\la\s^1_z\s^2_z\ra&=\la\s^1_z\ra\la\s^2_z\ra,
\end{align}
where $N\equiv N(\omega_0)$. The above solutions indicate equal population distributions of both molecules and the 
absence of intermolecular electronic coherence. 

In contrast, for $\Gamma\neq 0$, we obtain $\la\s^1_z\ra\neq \la\s^2_z\ra$, and a non-trivial two-molecule coherence in the two-level system.
Below we present the explicit expressions for two quantities: the excitation current, $\im\{\la\s^2_+\s^1_-\ra\}$, and the difference between the excited state populations of the two molecules, $\la\s^2_+\s^2_-\ra\! -\!\la\s^1_+\s^1_-\ra=(\la\s^2_z\ra\!-\!\la\s^1_z\ra)/2$:
\be
\im\{\la\s^1_-\s^2_+\ra\}\!=\!\frac{2N\Gamma[(1\!+\!2N)\gamma_1\gamma_2\Delta\!+\!(\gamma_2\!-\!\gamma_1)\Gamma\Omega]}{R} \label{exc_cur},
\e
\be
\la\s^2_+\s^2_-\ra\!-\!\la\s^1_+\s^1_-\ra\!=\!\frac{2N\Gamma(\gamma_1\!+\!\gamma_2)
[\Delta\Omega\!+\!(\gamma_1\!-\!\gamma_2)(1\!+\!2N)\Gamma]}{R}\label{diff_pop},
\e
with
\begin{align}
 R&\!=\!2(1\!+\!2N)(\gamma_2\!-\!\gamma_1)\Delta \Gamma\Omega\!+\!\Gamma^2[(1+2N)^2\n\\
 &\!\times\!\{2N(\gamma_1\!-\!\gamma_2)^2\!-\!(\gamma_1\!+\!\gamma_2)^2\}\!-\!4\Omega^2]\n\\
 &\!+\!(1+2N)^3[\gamma_1\gamma_2\{(1+2N)^2(\gamma_1\!+\!\gamma_2)^2\!+\!\Delta^2\}\n\\
 &\!+\!(\gamma_1\!+\!\gamma_2)^2\Omega^2].
 \label{R}
 \end{align}
Direct inspection of Eqs. (\ref{exc_cur}) and (\ref{diff_pop}), 
for $\gamma=\gamma_1=\gamma_2$, yields the energy balance relation (\ref{en_bal}).

\section{Monte-Carlo simulation of the stochastic current}
\label{sec:mc}
As shown in \cite{carmichael,dalibard92}, a density operator $\varrho(t)$ obeying a Markov master equation with a relaxation in Lindblad form can be unravelled into an ensemble of stochastic wavefunctions (quantum trajectories) $|\psi(t)\rangle$, such that averaging over possible outcomes at time $t$ yields the density operator, i.e.,
\be
\overline{|\psi(t)\rangle\langle \psi(t)|}=\varrho(t).
\label{average}
\e
Quantum trajectories corresponding to master equations with a unique steady state possess the property of ergodicity \cite{maassen93}. Therefore, when one deals with a steady state density matrix, it is more convenient to use the time average over a single trajectory instead of the ensemble average \cite{gardiner}.  

The master equation \eqref{meq} is not given in Lindblad form, but can be brought into it by a unitary transformation applied to the second line of \eqref{meq}. Then we obtain
\be
\dot{\varrho}=-(i/\hbar)[H_D,\varrho]+{\cal L}(\varrho),
\label{meq2}
\e 
where $H_D$ is given in Appendix \ref{app1}, 
and
\be
{\cal L}(\varrho)=\sum_{k}^4\lt(A_k\varrho A^\dagger_k-\frac{1}{2}\lt\{A^\dagger_kA_k,\varrho\rt\}\rt),
\e
with $A_1=\sqrt{2\gamma N}\sigma^1_+$, $A_2=\sqrt{2\gamma N}\sigma^2_+$, $A_3=\sqrt{\gamma(1+N)-\Gamma}(\sigma^2_--\sigma^1_-)$, $A_4=\sqrt{\gamma(1+N)+\Gamma}(\sigma^2_-+\sigma^1_-)$. We note that the contributions due to the Lindblad operators $A_1$, $A_2$, are $A_3$ are much smaller than the one due to $A_4$ (since $N=0.01$ and we consider the intermolecular distances such that $\Gamma\approx \gamma$).

First, we define the effective non-Hermitian Hamiltonian
\be
H_{eff}=H_D-(i\hbar/2)\sum_kA^\dagger_kA_k.
\label{Heff}
\e
Using Eq. \eqref{meq2}, \eqref{Heff}, we perform a stochastic unravelling as described in \cite{moelmer93}. We assume that the molecular aggregate is initially in its ground state, that is, $|\psi(0)\rangle=|g_1,g_2\rangle$, and 
divide the time axis into infinitesimal intervals $\delta t$ which should be much shorter than the shortest characteristic system time scale 
defined by $\Omega$. To generate the 
exemplary trajectory of the stochastic current in Fig. \ref{traj}, we fix 
$\delta t=2.0\times 10^{-5}\gamma^{-1}$ and $\Omega^{-1}\approx 8.4\times 10^{-5}\gamma^{-1}$. At each time step, we calculate the probability 
\be
\delta p=i\delta t\langle \psi(t)|H_{eff}-H_{eff}^\dagger|\psi(t)\rangle/\hbar,
\e
that the system evolves continuously, and update the quantum state as follows: We compared $\delta p$ with a random number $\epsilon$ uniformly distributed on the interval $[0,1]$. If $\delta p<\epsilon$, then $|\psi(t+\delta t)\rangle$ is given by
\be
|\psi(t+\delta t)\rangle=(1-(i/\hbar H_{eff})\delta t)|\psi(t)\ra/(1-\delta p)^{1/2}.
\e
At this stage, the state continuously evolves, undergoing coherent oscillations at the frequency $\Omega^\prime$ given by the eigenvalues of $H_D$ (see above). If $\delta p\geq \epsilon$, then a ``quantum jump'' occurs, whereby the state changes according to 
\be
|\psi(t+\delta t)\rangle=A_m|\psi(t)\rangle/(\delta p_m/\delta t)^{1/2}, \quad (m=1,\ldots, 4)
\e
with 
\be
\delta p_m=\delta t\la \psi(t)|A^\dagger_mA_m|\psi(t)\ra, \quad \sum_m \delta p_m=\delta p.
\e
It follows from the definitions of the jump operators $A_m$ that if the aggregate is in its ground state, a photoabsorption can be mediated by either $A_1$ or $A_2$, with the equal probability of 1/2. Once a photon is absorbed at a random time $t_0>0$, the probability $p_4$ of the photoemission associated with the operator $A_4$ is much larger than $p_3$, associated with $A_3$ 
(see above). Furthermore, because $N\ll 1$, the probability of double excitation is very low. However, if the next quantum jump does not occur until times $t-t_0\gtrsim 10\gamma^{-1}$, the dimer is effectively driven into the antisymmetric state $\propto |e_1,g_2\rangle-|g_1,e_2\rangle$, wherefrom it can undergo a jump either into the ground state mediated by the operator $A_3$, with the probability of $\approx 1/2$, or into the doubly excited state mediated by the operators $A_1$ or $A_2$ (each event with approximately equal probability of $\approx 1/4$).

To show why the dimer's state conditioned by the absence of a quantum jump following a photoabsorption becomes, at long times, the antisymmetic state, we turn to the non-unitary evolution operator generated by $H_{eff}$. This operator reads (in the basis $\{|g_1,g_2\rangle, |e_1,g_2\rangle, |g_1,e_2\rangle, |e_1,e_2\rangle\}$):
\begin{widetext}
\begin{equation*}
e^{-\frac{i H_{eff}t}{\hbar}}\!\!=\!\!\left(
\begin{array}{cccc}
e^{-2N\gamma t}&0&0&0\\
0&e^{-(1+2 N)\gamma t}\!\!\left[\cosh(\frac{\Omega'' t}{2})-\frac{i\Delta\sinh(\frac{\Omega'' t}{2})}{\Omega''})\right]&-e^{-(1+2 N)\gamma t}\frac{2(\Gamma+i\Omega)\sinh(\frac{\Omega'' t}{2})}{\Omega''}&0\\
0&-e^{-(1+2 N)\gamma t}\frac{2(\Gamma+i\Omega)\sinh(\frac{\Omega'' t}{2})}{\Omega''}&e^{-(1+2 N)\gamma t}\!\!\left[\cosh(\frac{\Omega'' t}{2})+\frac{i\Delta\sinh(\frac{\Omega'' t}{2})}{\Omega''})\right]&0\\
0&0&0&e^{-2(1+N)\gamma t}
\end{array}
\right),\!
\end{equation*}  
\end{widetext} 
where $\Omega''=\sqrt{4\Gamma^2-\Delta^2-4\Omega^2+8i\Gamma\Omega}$ is a complex frequency with $\im{\Omega''}\approx \Omega^\prime$ (see Appendix \ref{app1}) and $0<\re{\Omega''}<2\gamma$. Let us assume that at $t_0>0$ the dimer jumps into state $|e_1,g_2\ra$ via photoabsorption by molecule 1. At short times $(t-t_0)\ll \gamma^{-1}$, $\exp(-i H_{eff}(t-t_0)/\hbar)\approx \exp(-iH_D(t-t_0)/\hbar)$, such that $H_{eff}$ generates oscillations at the frequency $\Omega^\prime$ of the probability amplitudes associated with the states $|e_1,g_2\rangle$ and $|g_1,e_2\rangle$ (coherent excitation exchange between the molecules). At long times $(t-t_0)\gtrsim 10\gamma^{-1}$, the dominant contribution to the relaxation part of $H_{eff}$ is given by the operator $-(i\hbar/2)A^\dagger_4A_4\propto (\sigma_+^1+\sigma_+^2)(\sigma_-^1+\sigma_-^2)$. Consequently, the symmetric superpositions $|e_1,g_2\rangle+|g_1,e_2\rangle$, also known as superradiant states \cite{mandel95,ficek02}, decay faster than the antisymmetric (subradiant) states $|e_1,g_2\rangle-|g_1,e_2\rangle$ and
the conditioned state is given by
\begin{align}
|\psi(t-t_0)\ra_c&=\exp[\{\Omega''/2-(1+2N)\gamma\}(t-t_0)]\n\\
&\times(a|e_1,g_2\ra+b|g_1,e_2\ra),
\label{decaying_state}
\end{align} 
where $a=1/2-i\Delta/(2\Omega'')$ and $b=-(\Gamma+i\Omega)/\Omega''\approx -a$. Hence, $|\psi(t-t_0)\ra_c\propto |e_1,g_2\ra-|g_1,e_2\ra$.

Given a normalized state $|\psi(t)\ra$, we 
determine the stochastic excitation current by
\be
I_{\rm stoch}(t)=2\Omega \im\{\la \psi(t)|\sigma^2_+\sigma^1_-|\psi(t)\ra\}.
\label{stoch_curr}
\e
In particular, normalizing the state \eqref{decaying_state}, we obtain that the corresponding conditioned excitation current is time-independent:
\be
I_c= \frac{2\Omega\im(a b^*)}{|a|^2+|b|^2},
\e
and, for the parameters chosen in Fig.~\ref{traj}, $I_c\approx 0.0021 \gamma^{-1}$.

By virtue of Eq. \eqref{average}, the average over the ensemble of random realisations of $|\psi(t)\ra$ in \eqref{stoch_curr} yields the average excitation current as given by Eq. (\ref{exc_cur0}). On the other hand, the average steady state current can be obtained by the time average over a single quantum trajectory,
\be
\bar{I}_{\rm stoch}(\infty)=\lim_{T\to\infty}\frac{1}{T}\int_0^T{\rm d}t\; I_{\rm stoch}(t).
\e

\bibliography{thermal_coherence}

\end{document}